\documentclass{icrc2009}

\usepackage{graphicx}   
\usepackage[font=footnotesize]{subfig} 
\usepackage{fixltx2e}
\usepackage{url}

\newcommand{\shorttitle}[1]%
{\markboth{Proceedings of the 31\MakeLowercase{$^{st}$} ICRC, {\L}\'{o}d\'{z} 2009}{#1} }
\newcommand{\etal}{\MakeLowercase{\textit{et al. }}} 

\begin{document}
\title{A Monte Carlo study to check the hadronic interaction models by a new EAS hybrid experiment in Tibet}

\author{\IEEEauthorblockN{Ying Zhang\IEEEauthorrefmark{1}   \IEEEauthorrefmark{2},
J. Huang\IEEEauthorrefmark{1},
L. Jiang\IEEEauthorrefmark{1}   \IEEEauthorrefmark{3},
D. Chen\IEEEauthorrefmark{4},
L.K. Ding\IEEEauthorrefmark{1},
M. Shibata\IEEEauthorrefmark{5},\\
Y. Katayose\IEEEauthorrefmark{5},
N. Hotta\IEEEauthorrefmark{6},
M. Ohnishi\IEEEauthorrefmark{4},
T. Ouchi\IEEEauthorrefmark{7}and
T. Saito\IEEEauthorrefmark{8}}
\\ 
\IEEEauthorblockA{\IEEEauthorrefmark{1}Institute of High Energy Physics, Chinese Academy of Sciences, Beijing 100049, China}
\IEEEauthorblockA{\IEEEauthorrefmark{2}Institute of Modern Physics, Southwest Jiaotong University, Chengdu 610031, China}
\IEEEauthorblockA{\IEEEauthorrefmark{3}Department of Physics, Yunnan University, Kunming 650091, China}
\IEEEauthorblockA{\IEEEauthorrefmark{4}Institute for Cosmic Ray Research, University of Tokyo, Kashiwa 277-8582, Japan}
\IEEEauthorblockA{\IEEEauthorrefmark{5}Faculty of Engineering, Yokohama National University, Yokohama 240-8501, Japan}
\IEEEauthorblockA{\IEEEauthorrefmark{6}Faculty of Education, Utsunomiya University, Utsunomiya 321-8505, Japan}
\IEEEauthorblockA{\IEEEauthorrefmark{7}Faculty of Engineering, Kanagawa University, Yokohama 221-8686, Japan}
\IEEEauthorblockA{\IEEEauthorrefmark{8}Tokyo Metropolitan College of Industrial Technology, Tokyo 116-8523, Japan} 

}

\shorttitle{Huang \etal check hadronic interaction models}
\maketitle

\begin{abstract}
	A new EAS hybrid experiment has been designed by constructing a YAC (Yangbajing Air shower Core) detector 
	array inside the existing Tibet-III air shower array. The first step of YAC, called "YAC-I", consists
	of 16 plastic scintillator units (4 rows times 4 columns) each with an area of 40 cm$\times$50 cm
	which is used to check hadronic interaction models used in AS simulations. A Monte Carlo study 
	shows that YAC-I can record high energy electromagnetic component in the core region of air showers 
	induced by primary particles of several tens TeV energies where the primary composition is directly 
	measured by space experiments. It may provide a direct check of the hadronic interaction models currently 
	used in the air shower simulations in the corresponding energy region. In present paper, the method of the 
	observation and the sensitivity of the characteristics of the observed events to the different 
	interaction models are discussed.\\ 
\end{abstract}

\begin{IEEEkeywords}
	cosmic ray, hadronic interaction, Extensive Air Shower\\
\end{IEEEkeywords}

\section{Introduction}
The study of primary cosmic-ray energy spectrum and composition above *100 TeV has to depend on 
the indirect observation of extensive air showers (EASs) because of their low fluxes and the limited 
detetor acceptance of the on board satellite or baloon experiments. To interprete the EAS data, 
Monte Carlo (MC) simulations are inevitable. Any hadronic interaction models used in Monte Carlo 
codes are based on the knowledge obtained from the accelerator hadron-nucleus collision experiments. 
For accelerator experiments with energies lower than 2 TeV -- the corresponding fixed-target energy 
of the highest ISR energy, the inelastic interaction cross section, the interaction inelasticity and 
the distribution of large x (Feynman variable) particles (or particles produced in the forward region) 
have essentially been measured. However, when energy goes higher the inelasticity and the distribution 
of large x particles produced were no longer directly measured by hadron collider experiments, and 
one has to use extrapolation of the laws established in lower energies. Obviously, the correctness of 
the extrapolation determines the correctness of the description of EASs in higher energies. Nowadays 
many Monte Carlo simulations, when using different interaction models, resulted in different conclusions 
of the cosmic ray composition at the knee[1]. For multi-parameter measurements of EASs it seemed that 
no one interaction model can explain all data consistently. This situation asked for a further check 
and an improvement of the currently used interaction models.\\ 

Now we are going to have a chance to know more on the interaction features in the very forward region 
from the LHC collider experiments such as LHCf, TOTEM and CASTOR[2]. However, since the corresponding 
energy in the laboratory system of LHC reaches $10^{17}$eV, for some physics quantities or features, if they 
change with energy, one still needs to know how they are in $10^{14}$eV, $10^{15}$eV, $10^{16}$eV, etc.
Here we propose an approach to check the hadronic interaction models by observing EAS cores at an energy 
of *10 TeV using the newly constructed AS core detectors YAC-I.

\section{YAC-I experiment}
We have planed a new EAS hybrid experiment called YAC (Yangbajing Air shower Core array) in Tibet, 4300 m 
above sea level (an atmosphere depth of 606 g/cm$^{-2}$) aiming at the measurement of the primary p, He and 
Fe spectra in the knee region. Its first phase called YAC-I that consists of 16 EAS core detectors (as shown in Fig.1)
has been constructed and started data taking since April, 2009. YAC-I is located near the center of the Tibet-III 
air-shower array[3], operating simultaneously with Tibet-III and surrounding burst detectors (BD)[1]. For the
coincident events Tibet-III provides the total energy and the direction of air showers. BD can provide more 
accurate information about the core position and YAC-I observes high energy electromagnetic particles in the 
core region.\\

The 16 plastic scintillator units of YAC-I are arranged as an array (4 rows times 4 columns) each with an 
area of 40 cm$\times$50 cm. For our purpose, all detector units should be placed as densely as possible. 
A 28 cm spacing along x-axis and a 18 cm spacing along y-axis between two neighboring detectors are taken.
Each detector consists of lead plates with a thickness of 3.5 cm above the scintillator to convert high
energy electrons and gammas to showers. Each unit of YAC-I is attached with two photomultipliers to cover 
the wide dynamic range from 1 MIP (Minimum Ionization Particle) to 10$^{6}$ MIPs that corresponds to
*10 TeV of electron or gamma energy[4,5].

\begin{figure}[!t]
	\centering
	\includegraphics[width=3.1in]{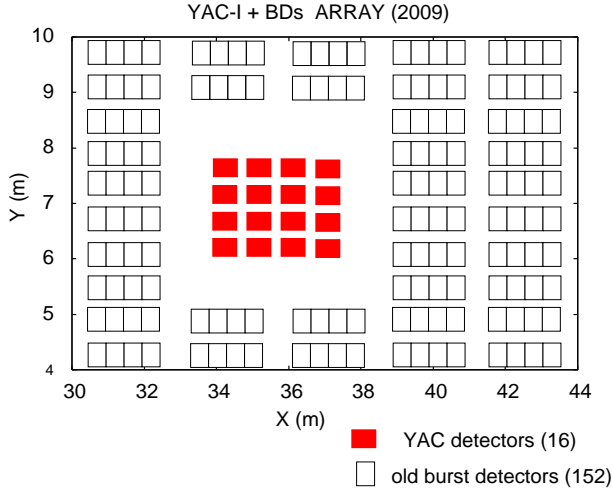}
	\caption{YAC-I + BDs ARRAY}
	\label{simp_fig}
\end{figure}

\section{Simulations and Analysis }
A Monte Carlo simulation has been carried out on the development of EAS in the atmosphere and the response 
in YAC-I. The simulation code CORSIKA (version 6.024) including QGSJET01c and SIBYLL2.1 hadronic interaction models[6]
are used to generate AS events.
Primary composition is taken from JACEE and RUNJOB[7,8]. At around several TeV to several 10 TeV
region, primary composition has been better measured. Primaries isotropically incident at the top of the 
atmosphere within the zenith angles from 0 to 60 degrees are injected into the atmosphere. The minimum 
primary energy is set 
at 1 TeV. Because of the use of Pb plates, secondary particles with lower energies could not reach at the 
scintillator. Thus, secondary particles are traced to the altitude of 4300 m till 300 MeV. The Monte Carlo 
air-shower events are randomly dropped onto the detector array plane, 15 m wider in each side of the 
rectangular-shape array. We choose the value 15 m because the area of 32.84 m$\times$32.14 m is checked to be
wide enough to contain 99.5\% EAS events under our event selection conditions (see below in the text). The electromagnetic 
showers in the lead 
layer induced by electrons or photons that hit any detector unit of the array are treated by a subroutine 
that is based on the detector simulation code EPICS[9]. The energy, position and angle of incident particle
and the structure of surrounding materials were taken into account in the detector simulation. Normally,
the following quantities of YAC are used to characterize an EAS core event:\\

\noindent 
${N_b}$ -- the number of shower particles hitting a detector unit;\\
${N_d}$ -- number of 'fired' detector units each with ${N_b}$$\geq$ a given threshold value;\\
${N_b}$$^{top}$ -- the maximum burst size among fired detectors;\\
$\sum$${N_b}$ -- total burst size of all fired detector units;\\
$<$R$>$ -- the mean lateral distance from the air shower core to a detector unit.\\

By using different threshold of ${N_b}$, different ${N_d}$, different ${N_b}$$^{top}$, etc, one can obtain
different event samples that have different average primary energy and different sample size. For various
physics goals one may use different event selection criteria. To see how some physics quantities change
with energy one may use different samples simultaneously. Now, only for an example, we choose a sample at the
energies of several 10 TeV region. This event sample has been selected under the following conditions:\\

\noindent
(1) the number of shower particles hitting a detector unit ${N_b}$$\geq$100;\\
(2) the number of fired detectors ${N_d}$$\geq$4;\\
(3) the detector unit with ${N_b}$$^{top}$ is located at the inner 4 detectors of YAC-I grid in order to reject 
events falling far from the array.  
\begin{table}[!h]
	\caption{The fractions of the components after the burst event selection}
	\label{table1}
	\centering
	\begin{tabular}{|c|c|c|c|}
		\hline
		Int. Model &Component&  ${E_{low}}$(TeV)\footnotemark  &${E_{high}}$(TeV) \\
		\hline
		&         &1-20 & 20-200\\ 
		\cline{2-4}  
		& proton       &96.1 & 73.6\\
		& He           &3.8  & 21.8 \\
		QGSJET   & Medium(CNO)  &0.1  & 2.9 \\
		& Heavy(NaMgSi)   &0 & 1.2 \\
		& Very Heavy(SClAr)  &0  & 0.2 \\
		& Fe      &0  & 0.3 \\
		\hline
		& proton       &94.9 & 70.9\\
		& He            &5.0  & 23.5 \\
		SIBYLL   & Medium(CNO)      &0.1  & 3.5 \\
		& Heavy(NaMgSi)      &0  & 1.3 \\
		& Very Heavy(SClAr)  &0  & 0.1 \\
		& Fe      &0  & 0.7 \\
		\hline
	\end{tabular}
\end{table}
\footnotetext {${E_{low}}$ indicates lower primary energy of the selected sample,${E_{high}}$ indicates higher primary energy of the selected sample.}\\
We sampled 7.28$\times$$10^{9}$ and 5.41$\times$$10^{9}$ primaries for the QGSJET and SIBYLL model, 
respectively. After the event selection 18715 and 17166 burst events passed for the QGSJET and SIBYLL model, 
respectively. The average generation efficiencies of the burst events in this energy by SIBYLL is higher 
than QGSJET by a factor of 1.24. The attenuation length ($\lambda$) of the burst events for the QGSJET and 
SIBYLL model is estimated by using the zenith angle distribution which are 109 $\pm$ 4 (g/cm$^{-2}$) and 
108 $\pm$ 4 (g/cm$^{-2}$), respectively. It is an important parameter reflecting the characteristics of 
inelastic cross section and inelasticity used in the model.\\

It is seen that after the event selection the left burst events are mostly (about $\sim$96\%) induced by 
protons and heliums. The fractions of the components of the burst events are summarized in Table 1. This 
is suitable for our aims because primary proton and helium spectra were better measured than other heavier 
nuclei[7,8] and the systematic uncertainty induced by other nuclei will be smaller than 4\%. In order to 
observe interaction model dependences, the experimental data will be analyzed in the same manner as for the 
MC data in the procedure.\\

\section{Results and Discussion}

The primary-energy distribution of burst events for QGSJET and SIBYLL are obtained as shown in Fig.2. 
Fig.2 shows a peak at around 80 TeV for both models, just meeting our requirement. This is the energy 
region we are going to check the interaction models in the first step.\\ 

Fig.3 is the spectrum of the total burst size $\sum$Nb which should depend sensitively on the inelastic interaction 
cross section, the inelasticity, and particles produced in the forward region. The difference of the flux intensity 
between two interaction models is 1.26 $\pm$ 0.05.
Fig.3 shows an obvious difference of flux intensity between two models, and one may find the reason from the difference of the two models.
To compare YAC observation of the burst size flux with these MC predictions, we can provide some evidences on the feature
of above mentioned physics quantities.\\

\begin{figure}[!t]
	\centering
	\includegraphics[width=3.1in]{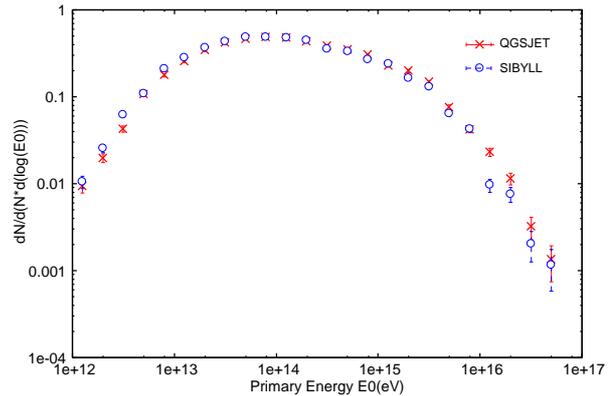}
	\caption{The distribution of primary energy of the sample selected for QGSJET model and SIBYLL model. 
	The peak position of the primary energy spectrum for both models is about 80 TeV, just meeting our requirement. }
	\label{simp_fig}
\end{figure}

\begin{figure}[!t]
	\centering
	\includegraphics[width=3.1in]{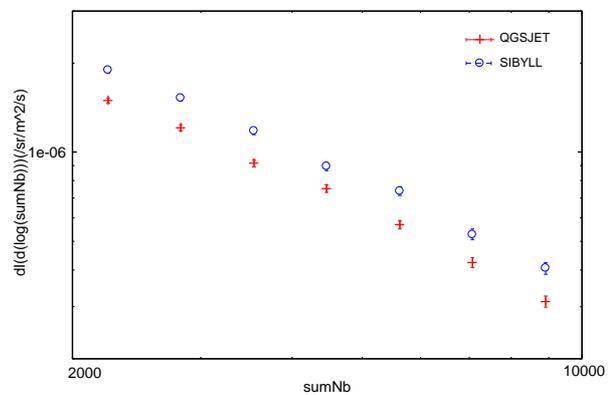}
	\caption{The total burst size ($\sum$Nb) spectrum obtained by QGSJET and SIBYLL model, where sumNb indicates the 
	total burst size ($\sum$Nb).}
	\label{simp_fig}
\end{figure}

\begin{figure}[!t]
	\centering
	\includegraphics[width=3.1in]{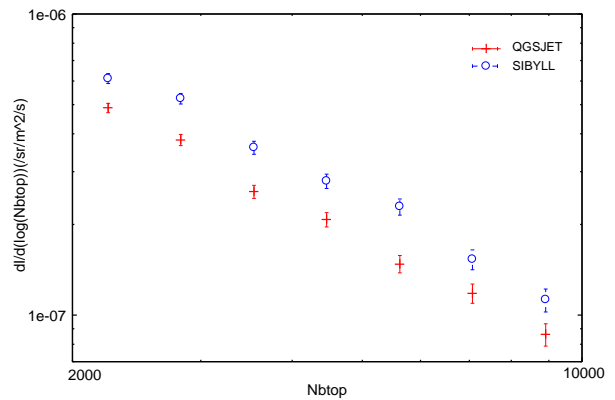}
	\caption{Top burst size (${N_b}$$^{top}$) spectrum obtained by QGSJET and SIBYLL model.}
	\label{simp_fig}
\end{figure}

The spectrum of the top burst size ${N_b}$$^{top}$ is obtained as shown in Fig.4.
It is found that there is a difference about 1.36 $\pm $ 0.05 in the slopes of two models in the energy region.
The quantity of ${N_b}$$^{top}$ may relate with
the features of leading particles and transverse momentum of secondary particles produced in hadronic interactions.\\

The ${N_d}$ distribution is shown in Fig.5, suggesting a visible model dependence between QGSJET and SIBYLL.
It reflects lateral characteristics of high energy particles in the air shower cores. Comparing with YAC observation, we can check 
these different hadronic interaction models. \\
\begin{figure}[!t]
	\centering
	\includegraphics[width=3.1in]{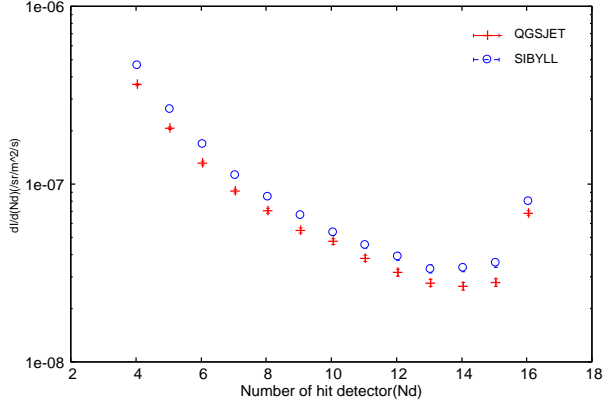}
	\caption{The spectrum of the number of hit detector (${N_d}$) obtained by QGSJET and SIBYLL model.}
	\label{simp_fig}
\end{figure}
\begin{figure}[!t]
	\centering
	\includegraphics[width=3.2in]{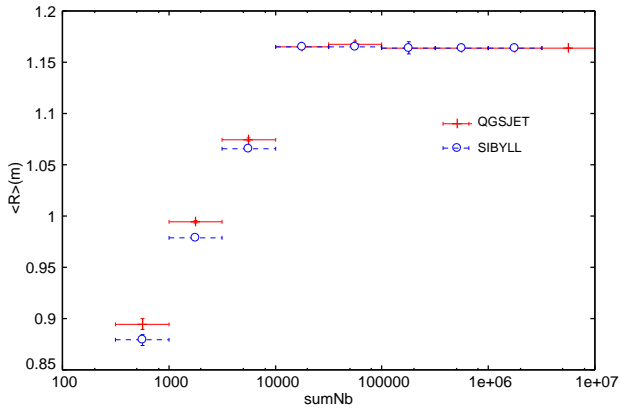}
	\caption{Distribution of the mean lateral spread ($<$R$>$) obtained by QGSJET and SIBYLL model, where 
	sumNb indicates the total burst size $\sum$Nb, $<$R$>$ indicates mean lateral spread.}
	\label{simp_fig}
\end{figure}

Fig.6 shows the mean lateral spread ($<$R$>$ = $\sum$${r_i}$/(${N_d}$-1)), where ${r_i}$ and ${N_d}$ are the 
lateral distance from the air shower core to the center of a fired detector and the number of hit detectors,
respectively. Here we have imposed the condition of ${N_d}$$\geq$4. The mean\\
\\
lateral spread $<$R$>$ will be changed
at different energy regions that can be reflected by total burst size $\sum$ $N_b$, as shown in Fig.6. It provides a check 
on the features of transverse momentum in the very forward region.\\ 

The area S of the event-dropping is 1.06 $\times$ $10^{3}$ m$^{2}$.
The effective solid angle is 2.355 sr. 
Taking 6 months as the effective observation time, the number of primary cosmic rays is calculated to be
7.74$\times$ $10^{9}$ particles, a
\newpage
\noindent
factor of 1.06 higher than the Monte Carlo sample. That is to say,
the experiment will produce enough statistics in 6 months.\\
\section{Conclusions}
The Monte Carlo shows that:\\
(1) Under the above selection conditions a sample of events with a primary energies at around 80 TeV can be 
selected. This is an energy region the primary composition being better measured directly;\\
(2) A sample with large statistics can be obtained in a few months's observation.\\

In summary, taking the priority of high altitude (like Yangbajing) an EAS core event sample at the energy region
around several 10 TeV can be obtained with high statistics by using YAC-I. The hadronic interaction models 
at this energy region can be checked. Emulsion chamber[1,10] cannot work at this lower energies. Other measurements 
at low altitudes are also difficult to target the similar aim of checking the hadronic interaction models.\\

\section{Acknowledgment}
The authors would like to express their thanks to the members of the Tibet AS$\gamma$ collaboration for the 
fruitful discussion.The support of the Knowledge Innovation Fund (H8515530U1) IHEP, China,to J. Huang is also acknowledged.\\

\end{document}